\documentclass[%
 reprint,
 /amsmath,amssymb,
 aps,
]{revtex4-1}

\usepackage[dvipdfmx]{graphicx}
\usepackage{dcolumn}
\usepackage{bm}
\usepackage{epstopdf}
\usepackage{epsfig}
\usepackage{natbib}
\usepackage{color}
\usepackage{amsmath}

\newcommand\etal{\mbox{\textit{et al.}}}
\def\tr{\textcolor{black}}
\def\trr{\textcolor{black}}
\def\tb{\textcolor{black}}
\def\tg{\textcolor{black}}

\begin{document}

\preprint{APS/123-QED}

\title{Effects of pressure impulse and peak pressure of a shock wave on microjet velocity and the onset of cavitation in a microchannel}

\author{Keisuke Hayasaka}

\author{Akihito Kiyama}
\author{Yoshiyuki Tagawa}%
 \email{tagawayo@cc.tuat.ac.jp} 
\affiliation{%
Department of Mechanical Systems Engineering, Tokyo University of Agriculture and Technology,
Naka-cho 2-24-16, Koganei, Tokyo 184-8588, Japan
}%

\date{\today}

\begin{abstract}
The development of needle-free injection systems utilizing high-speed microjets is of great importance to world healthcare.
It is thus crucial to control the microjets, which is often induced by underwater shock waves.
In this contribution from fluid-mechanics point of view, we experimentally \tb{investigate} the effect of a shock wave on the velocity of a free surface (microjet) and underwater cavitation \tb{onset} \tr{in a microchannel}, focusing on the pressure impulse and peak pressure of the shock wave.
The shock wave used had a non-spherically-symmetric \tr{peak pressure} distribution and a spherically symmetric \tr{pressure impulse} distribution \tr{[Tagawa \textit{et al., J. Fluid Mech.}, 2016, \textbf{808}, 5-18]}. 
First, we \tb{investigate} the effect of the shock wave on the jet velocity by installing a narrow tube and a hydrophone in different configurations in a large water tank, and measuring the shock wave pressure and the jet velocity simultaneously. 
The results \tb{suggest} that the jet velocity depends only on the pressure impulse of the shock wave. 
We then \tb{investigate} the effect of the shock wave on the cavitation \tg{onset} by taking measurements in an L-shaped microchannel. 
The results \tb{suggest} that the probability of cavitation \tg{onset} depends only on the peak pressure of the shock wave. 
In addition, the jet velocity \tb{varies} according to the presence or absence of cavitation. 
The above findings provide new insights for advancing a control method for high-speed microjets.
\end{abstract}

\pacs{Valid PACS appear here}
\maketitle


\section{Introduction}
\label{Sec:Intro}
Needle-free injection using a liquid jet is an attractive alternative to conventional hypodermic injection in terms of infectious disease prevention\cite{mitragotri2006current}. 
However, in existing liquid jet injection methods, the jet is diffuse\cite{schramm2004needle}, which causes problems including pain and difficulty in localized administration, and these problems are a factor impeding the widespread use of needle-free injections. 
Recently, the authors  have developed a method of generating a focused jet (diameter of jet tip: several $\mu$m)\cite{tagawa2012highly, tagawa2013needle} that has the potential to resolve the existing problems. 
A characteristic of this method is the use of a laser-induced shock wave, which when travelling toward a concave air-liquid interface in a microtube filled with liquid, generates a high-speed ($\sim$850 m/s) microjet.
\tr{Delrot $\etal$ demonstrated that this method can be applied to the inkjet printing system\cite{delrot2016inkjet}.}
\tr{Other groups have reported the jet generating devices which utilize the interaction between the shock \tg{wave} and \tg{the} air-liquid interface\cite{avila2015fast, berrospe2016continuous, koita2016experimental, lu2016microliquid}.}

The velocity of the jet is known to be determined by the ``pressure'' of the shock wave because the jet is generated as a result of interaction between the air-liquid interface and the shock wave. 
In calculations that assume unsteady compressible flow, Turangan\cite{turangan2013highly} showed that the jet velocity repeatedly increases and decreases as it develops until it reaches a maximum. 
This finding suggests that the instantaneous pressure of the shock wave arriving at the air-liquid interface affects the movement of the air-liquid interface, and the {\it peak pressure} of the shock wave can be considered to be a determinant of the jet velocity. 
Meanwhile, Peters $\etal$ performed numerical calculations assuming incompressible potential flow\cite{peters2013highly}.
They showed that jet velocity is proportional to the gradient of the {\it pressure impulse} generated between the bubble and the interface within the timescale of jet formation. 
This shows that, within the said timescale, the jet velocity can be described without regard to the behavior of the shock wave. 
This finding is also consistent with a mesoscale analysis of a jet using an impact (water-hammer pressure wave)\cite{antkowiak2007short}. 
As described above, the peak pressure and pressure impulse of the shock wave are conceivable as physical quantities relating to the jet velocity. 
However, because it is difficult to control both of these independently, the physical quantities that determine the velocity of a focused microjet have not yet been identified.

In addition, when a relatively strong shock wave is reflected at the air-liquid interface, it can cause cavitation\cite{ando2012homogeneous}. 
In this microjet as well, cavitation bubbles have been identified between the laser-induced bubble and the air-liquid interface\cite{kawamoto2016}.
When cavitation occurs, the mode of shock wave propagation changes, and this causes an increase in the velocity of the jet\cite{kiyama2016effects}. 
The importance of the pressure impulse during bubble growth inside a narrow tube has been suggested\cite{ory2000growth}. 
Meanwhile, conditions for cavitation are usually based on the magnitude of pressure fluctuation (tensile strength) represented by the liquid vapor pressure\cite{herbert2006cavitation}. 
Because of changes in the form of the shock wave and difficulties in making small-scale pressure measurements, there are no examples of the roles of pressure impulse and peak pressure being measured and regulated in the same system.

Therefore, in this study, we experimentally investigate the effect of a shock wave on \tr{microjet velocity} and cavitation \tg{onset}, focusing on the pressure impulse and peak pressure of the shock wave. 
To control the magnitude of the pressure impulse and the peak pressure, shock waves with non-spherically symmetric pressure distributions are generated\cite{vogel1996plasma, sankin2008focusing}. 
An interesting characteristic of shock waves of this kind is that they have a spherically symmetric pressure impulse distribution, despite having a non-spherically symmetric peak pressure distribution\cite{tagawa2016pressure}. 
As a result, the magnitude of the peak pressure can be independently controlled by changing the direction of the laser irradiation. 
In addition, by changing the laser energy, the magnitude of the pressure impulse can be independently controlled. 
In this study, we make use of this characteristic to measure the pressure and jet velocity in the directions parallel and perpendicular to the laser irradiation direction. In addition, we observe cavitation \tb{onset} at this time using captured images, and \trr{discuss how our results can contribute to improve controllability.}

The remainder of this paper is outlined as follows. 
Section \ref{Sec:Method} describes the two experimental setups used in this study. 
Next, Section \ref{Sec:Result_A} describes the effect of shock wave pressure on jet velocity in the water tank experiment. 
Section \ref{SubSec:A1} describes the measurement of the shock pressure waveform through the water tank experiment, and Section \ref{SubSec:A2} describes the effect of the peak pressure and pressure impulse on the jet velocity. 
Next, Section \ref{Sec:Result_B} describes the effect of the shock wave on cavitation in the experiment inside an L-shaped \tb{microchannel}. 
Section \ref{SubSec:B1} describes the probability of cavitation, and Section \ref{SubSec:B2} describes the effect of the presence or absence of cavitation on jet velocity. 
Finally, Section \ref{Sec:Con} summarizes this paper.

\section{Method}
\label{Sec:Method}

\begin{figure}[h!!]
\includegraphics[width=1\columnwidth]{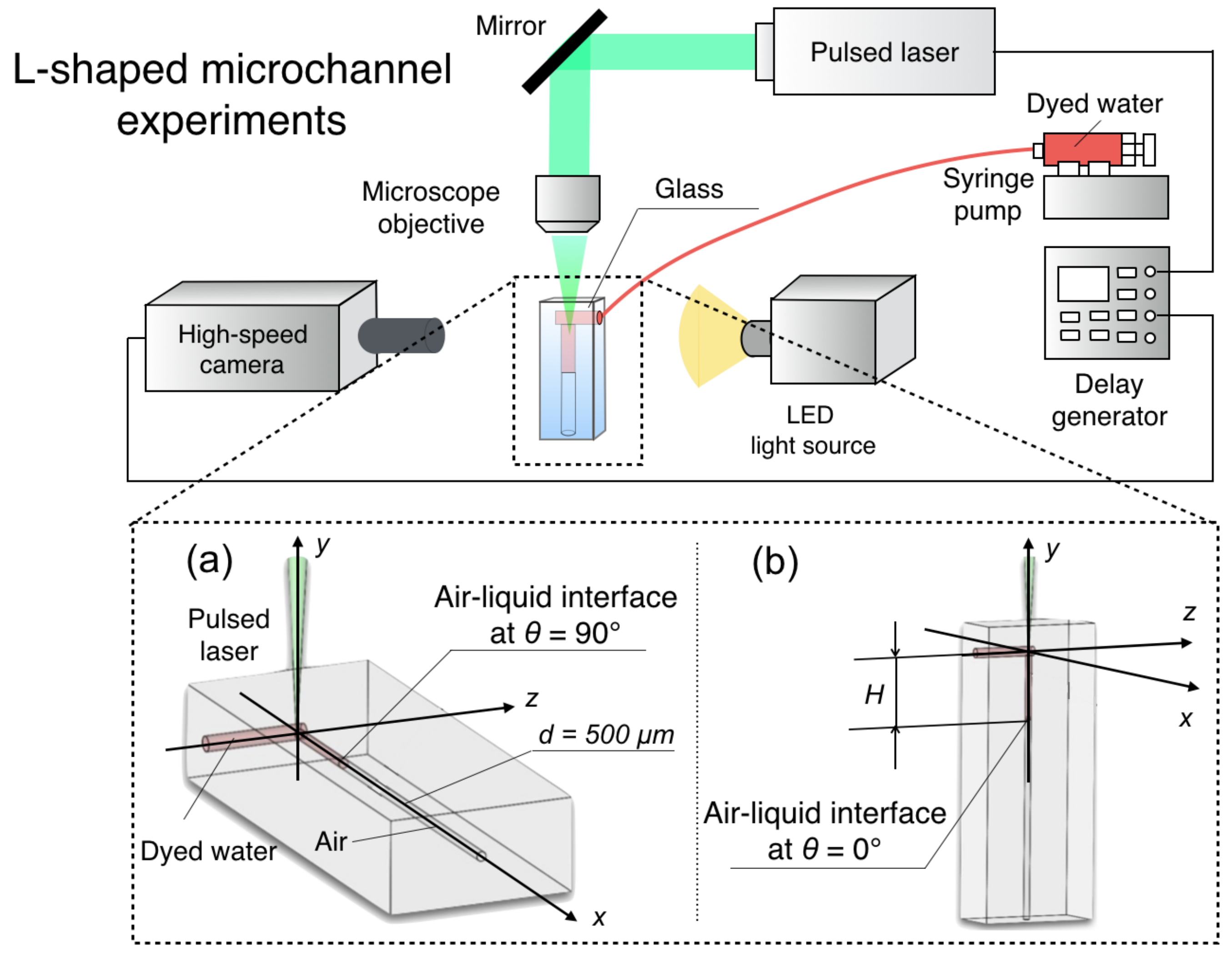}
\caption{
Experimental setup for generating microjets in L-shaped microchannel (L-shaped \tb{microchannel} experiments). 
(a)  and (b)  correspond to the microjet directed toward $\theta$ = 90$^\circ$ ($x$-axis) and $\theta$ = 0$^\circ$ ($y$-axis), respectively. 
The laser is incident from the positive side of the $y$-axis. 
}
\label{SetupB}       
\end{figure}

Two experimental setups were constructed for this study. 
The first setup was used to determine the effect of the shock wave pressure on the jet velocity in a water tank (Fig. \ref{SetupA}). 
The second setup was used to investigate the effect of the shock wave pressure on cavitation in an L-shaped \tb{microchannel} (Fig. \ref{SetupB}). 
Because it is difficult to measure the pressure in the microchannel, the pressure in the L-shaped \tb{microchannel} was assumed to be the same as the pressure measured in the water tank experiment.

\subsection{Simultaneous measurement of underwater shock wave pressure and microjet velocity (water tank experiments)}
\label{SubSec:Method_A}

\begin{table}[b]
  \begin{center}
    \caption{Variety of the microscope objective. N.A. : the numerical aperture, W.D. : the working distance, F.A. : the focusing angle of the objective lens, Diameter : the diameter of the focused laser beam.}
    \begin{tabular}{|c|c|c|c|c|} \hline
       Magnification & N.A & W.D. & F.A. & Diameter \\ \hline 
        5$\times$ & 0.10 & 20.0 mm & 1 $^\circ$ & 6.5 $\mu$m \\ \hline
      10$\times$ & 0.25 & 10.6 mm & 4 $^\circ$ & 2.6 $\mu$m \\ \hline
      20$\times$ & 0.25 & 25.0 mm & 6 $^\circ$ & 2.6 $\mu$m \\ \hline
    \end{tabular}
    \label{Table1}
  \end{center}
\end{table}

\begin{figure}[h!!]
\includegraphics[width=1\columnwidth]{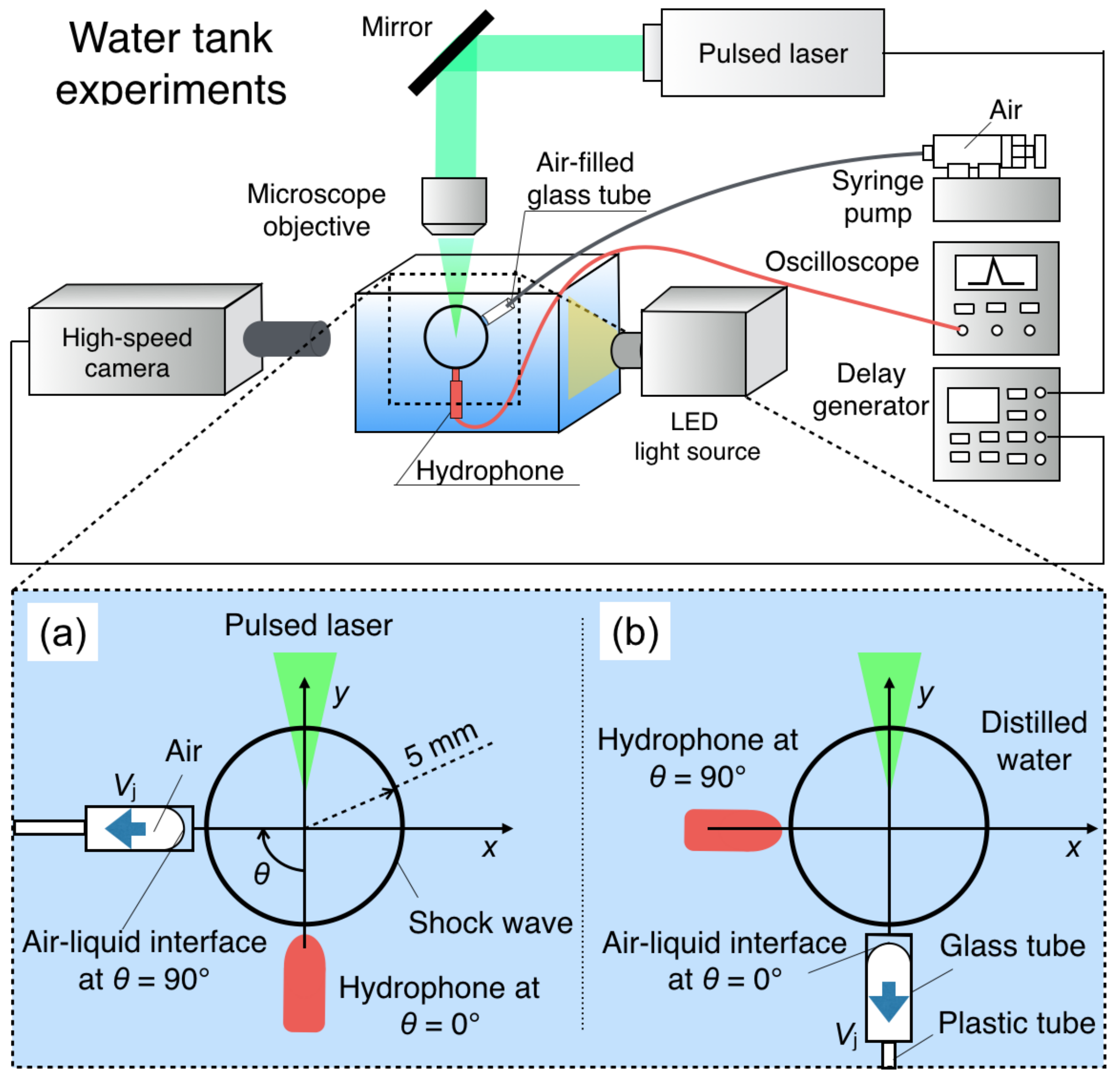}
\caption{
Experimental setup for measuring shock pressure and jet velocity (water tank experiments). 
Expanded views show the measurement area in the water tank. 
The laser spot is situated on the origin. 
(a) the hydrophone is placed on $\theta$ = 0$^\circ$ ($y$-axis) and the glass tube on $\theta$ = 90$^\circ$ ($x$-axis). (b) the hydrophone on $\theta$ = 90$^\circ$ ($x$-axis) and a glass tube on the $\theta$ = 0$^\circ$ ($y$-axis). 
}
\label{SetupA}       
\end{figure}

Fig.\ref{SetupA} shows the setup for measuring the underwater shock wave pressure and microjet velocity (hereafter, ``water tank experiments''). 
The method and conditions for generating shock waves are the same as those used in our previous study\cite{tagawa2016pressure}. 
An underwater shock wave is generated by a 532-nm pulsed laser (Nd:YAG; Nano S PIV, Litron Lasers) with a pulse duration of 6 ns, focused through an objective lens (MPLN series, \tb{Olympus}) inside a tank (300$\times$300$\times$450 mm$^3$) filled with water. 
The water is distilled using a water purification system (Milli-Q Integral, Merck; 13 M$\Omega$$\cdot$cm) at room temperature (15$\sim$20$^\circ$C) and gas saturated. 
Using three types of objective lenses makes it possible to change the magnitudes of the peak pressure and pressure impulse of the shock wave. 
Table \ref{Table1} shows the characteristics of each lens. 
The range of laser energy is 3-10 mJ, and the measurement error is $\pm$5$\%$.

The experiment is carried out using two setups (Fig.\ref{SetupA}(a) and (b)) in order to obtain the shock wave pressure and jet velocity in two directions. 
In Fig.\ref{SetupA}(a), hydrophone (Needle Probe, Mueller Instruments; measurable range: -10 - 100 MPa; rise time: 50 ns, measurement error: $\pm$2$\%$ \cite{zhou2016signal}) is installed along the laser irradiation direction (hereafter, $\theta$ = 0$^\circ$), and a narrow glass tube ($d$ = 500 $\mu$m) is installed perpendicular to the laser irradiation direction ($\theta$ = 90$^\circ$). 
Note that $\theta$ denotes the angle from the laser irradiation direction.
In Fig.\ref{SetupA}(b), the narrow glass tube is installed along the laser irradiation direction ($\theta$ = 0$^\circ$), and the hydrophone is installed perpendicular to the laser irradiation direction ($\theta$ = 90$^\circ$). 
The time history of the shock wave pressure is measured using an oscilloscope (ViewGo II DS-5554A, IWATSU; rise time: 750 ps; maximum sampling rate: 2 GS/s) connected to the hydrophone. 
The distance from the laser focal point to the end faces of the hydrophone and narrow glass tube is 5.0 $\pm$ 0.1 mm. 
One end of the narrow glass tube is connected via a plastic tube to an air-filled syringe. 
The position of the air-liquid interface in the glass tube is adjusted using a syringe pump (ULTRA 70-3005, Harvard). 
The other end of the glass tube is open. 
The position of the air-liquid interface is adjusted to be at the end of the tube. 
When a laser-induced shock wave arrives at the air-liquid interface, the air-liquid interface forms a jet directed toward the inside of the narrow tube.
The jet velocity $V_{j}$ is measured from images captured using a high-speed camera with a spatial resolution of 1,024$\times$124 pixels and a recording rate of 80,000 fps (FASTCAM SA-X, Photron). 
Specifically, $V_{j}$ is \tg{measured} between the image taken at the instant that the air-liquid interface inside the glass tube focuses and the next image, using the distance travelled by the tip of the jet. 
An LED light source (KL 1600 LED, OLYMPUS) is used. 
The pulsed laser and high-speed camera are synchronized using a delay generator (Model 575 Pulse/Delay Generator, BNC).

\subsection{Microjet formation in L-shaped microchannel\\(L-shaped \tb{microchannel} experiments)}
\label{SubSec:Method_B}

Fig.\ref{SetupB} shows the setup for forming a microjet inside the narrow tube (hereafter, ``L-shaped \tb{microchannel} experiments''). 
The equipment used to generate the jet and measure its velocity is the same as for the experiments in the water tank. 
The jet is formed inside the glass (quartz glass, 10$\times$5$\times$25 mm$^3$), which has an L-shaped \tb{microchannel}. 
One section of the \tb{microchannel} (800 $\mu$m in diameter) is connected to a syringe filled with dyed water. 
The energy absorption efficiency of dyed water is higher than that of pure water. 
Therefore, laser-induced bubbles can occur at lower laser energies than in the water tank experiments. 
The range of input energy of the laser is 102-625 $\mu$J, and the measurement error is $\pm$10$\%$. 
Only the 10$\times$objective lens is used. 
The position of the air-liquid interface is manipulated using a syringe pump. 
The range of distance $H$ from the air-liquid interface to the laser-induced bubble is 800-2700 $\mu$m. 
The other section of the \tb{microchannel} (diameter $d=$500 $\mu$m) is open to the atmosphere and contains the air-liquid interface. 
The pulsed laser is focused on the orthogonal portion of the \tb{microchannel}. 
In Fig.\ref{SetupB}(a), the air-liquid interface that generates the jet is perpendicular to the laser irradiation direction ($\theta$ = 90$^\circ$). 
In Fig.\ref{SetupB}(b), the air-liquid interface is in the laser irradiation direction ($\theta$ = 0$^\circ$). 
Because the inside diameter of the \tb{microchannel} is much smaller than the radius of the hydrophone, it is not possible to measure the pressure. 
In this experiment, the pressure distribution of the underwater shock wave generated in the experiments in the L-shaped \tb{microchannel} is assumed to be the same as the pressure distribution of the shock wave generated in the experiments in the water tank.

\section{Results and discussion\\(water tank experiments)}
\label{Sec:Result_A}

\subsection{Preliminary experiment}
\label{SubSec:A1}

\begin{figure}[t]
\includegraphics[width=1\columnwidth]{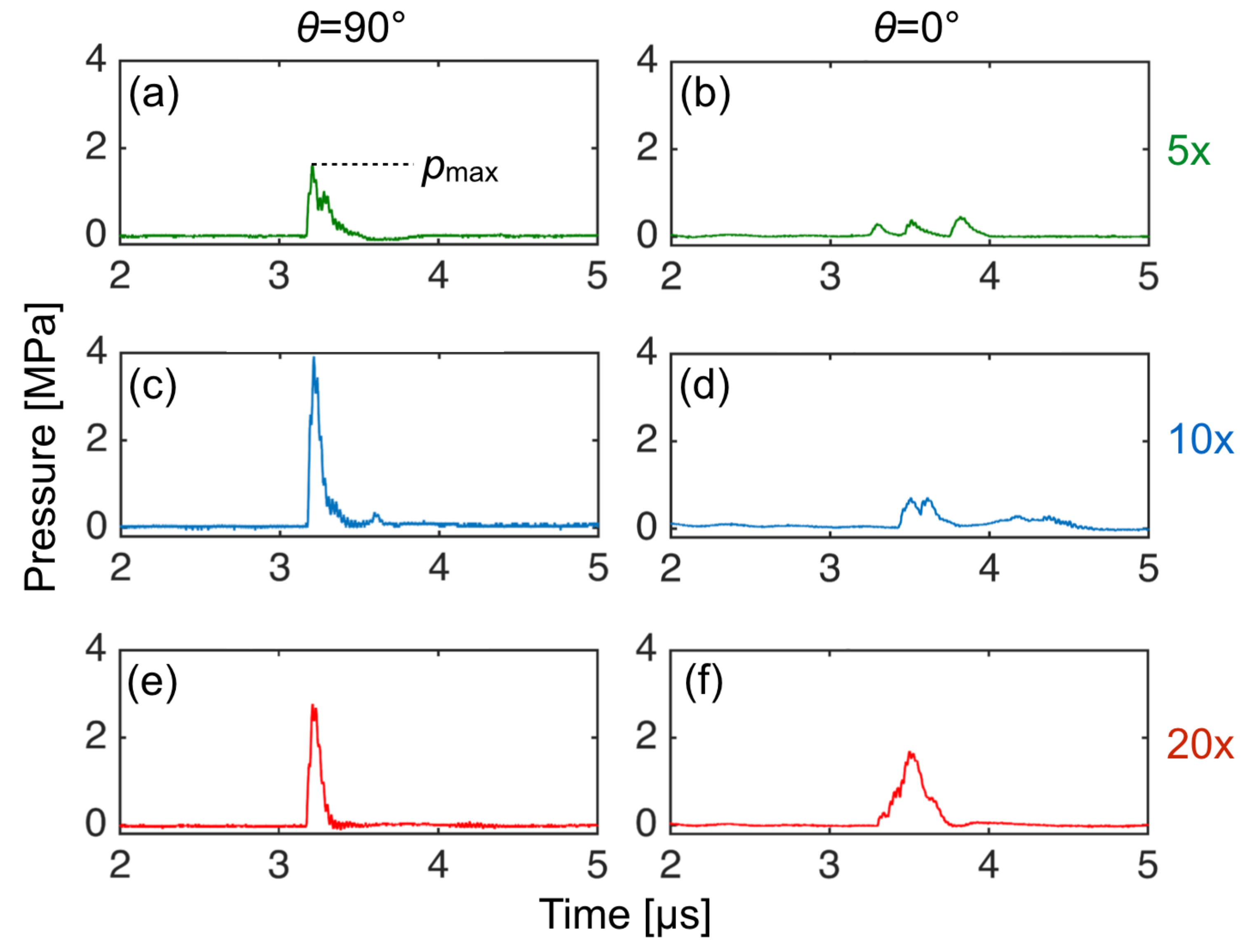}
\caption{
Pressure vs. time measured by hydrophones placed at $\theta$=90$^\circ$ and $\theta$=0$^\circ$. 
The horizontal and vertical axes represent the time from laser irradiation and the pressure detected by the hydrophone, respectively. 
(a) hydrophone: $\theta$=90$^\circ$, magnification: 5$\times$;
(b) hydrophone: $\theta$=0$^\circ$, magnification: 5$\times$; 
(c) hydrophone: $\theta$=90$^\circ$, magnification: 10$\times$; 
(d) hydrophone: $\theta$=0$^\circ$, magnification: 10$\times$;
(e) hydrophone: $\theta$=90$^\circ$, magnification: 20$\times$;
(f) hydrophone: $\theta$=0$^\circ$, magnification: 20$\times$. 
The laser energy is 3.0 mJ.
}
\label{Wave1}       
\end{figure}

This section confirms that the shock waves generated using these experimental setups are similar to those generated in our previous report\cite{tagawa2016pressure}. 
Fig.\ref{Wave1} shows time histories of shock wave pressure in the two directions measured in the water tank experiments. 
Note that the pressure in each direction is not measured simultaneously. 
The horizontal axis shows elapsed time from laser irradiation, and the vertical axis shows pressure. 
When the pressure is $p$ = 0, it indicates hydrostatic pressure. 
Fig.\ref{Wave1}(a) and Fig.\ref{Wave1}(b) show the results when a 5$\times$ objective lens is used, 
Fig.\ref{Wave1}(c) and Fig.\ref{Wave1}(d) show the results when a 10$\times$ objective lens is used, 
and Fig.\ref{Wave1}(e) and Fig.\ref{Wave1}(f) show the results when a 20$\times$ objective lens is used. 
Fig.\ref{Wave1}(a), Fig.\ref{Wave1}(c), and Fig.\ref{Wave1}(e) show the pressure waveform obtained for $\theta$ = 90$^\circ$, and  Fig.\ref{Wave1}(b), Fig.\ref{Wave1}(d), and Fig.\ref{Wave1}(f) show the pressure waveform obtained for $\theta$ = 0$^\circ$.
The laser energy was taken to be 3.0 $\pm$ 0.6 mJ. 
The pressure waveform obtained for $\theta$ = 90$^\circ$ differs greatly from that obtained for $\theta$ = 0$^\circ$. 
For $\theta$ = 90$^\circ$, there is one sharp pressure peak. 
For $\theta$ = 0$^\circ$, the pressure changes gradually, and there is a greater number of pressure peaks. 
This agrees well with our report\cite{tagawa2016pressure}, so the shock wave generated using these experimental setups is considered to have an non-spherically-symmetric pressure distribution, similar to that generated previously. 
In our report\cite{tagawa2016pressure}, we discussed the mechanism in detail and demonstrated that the peak pressure of the shock wave differs according to the propagation direction, but the pressure impulse is equal regardless of the propagation direction. 
In the experimental setups used in this study as well, the pressure impulse is considered to show similar tendencies to those reported in our report\cite{tagawa2016pressure}.

\subsection{Effect of pressure impulse and peak pressure on \tr{microjet velocity}}
\label{SubSec:A2}

\begin{figure}[t]
\includegraphics[width=0.95\columnwidth]{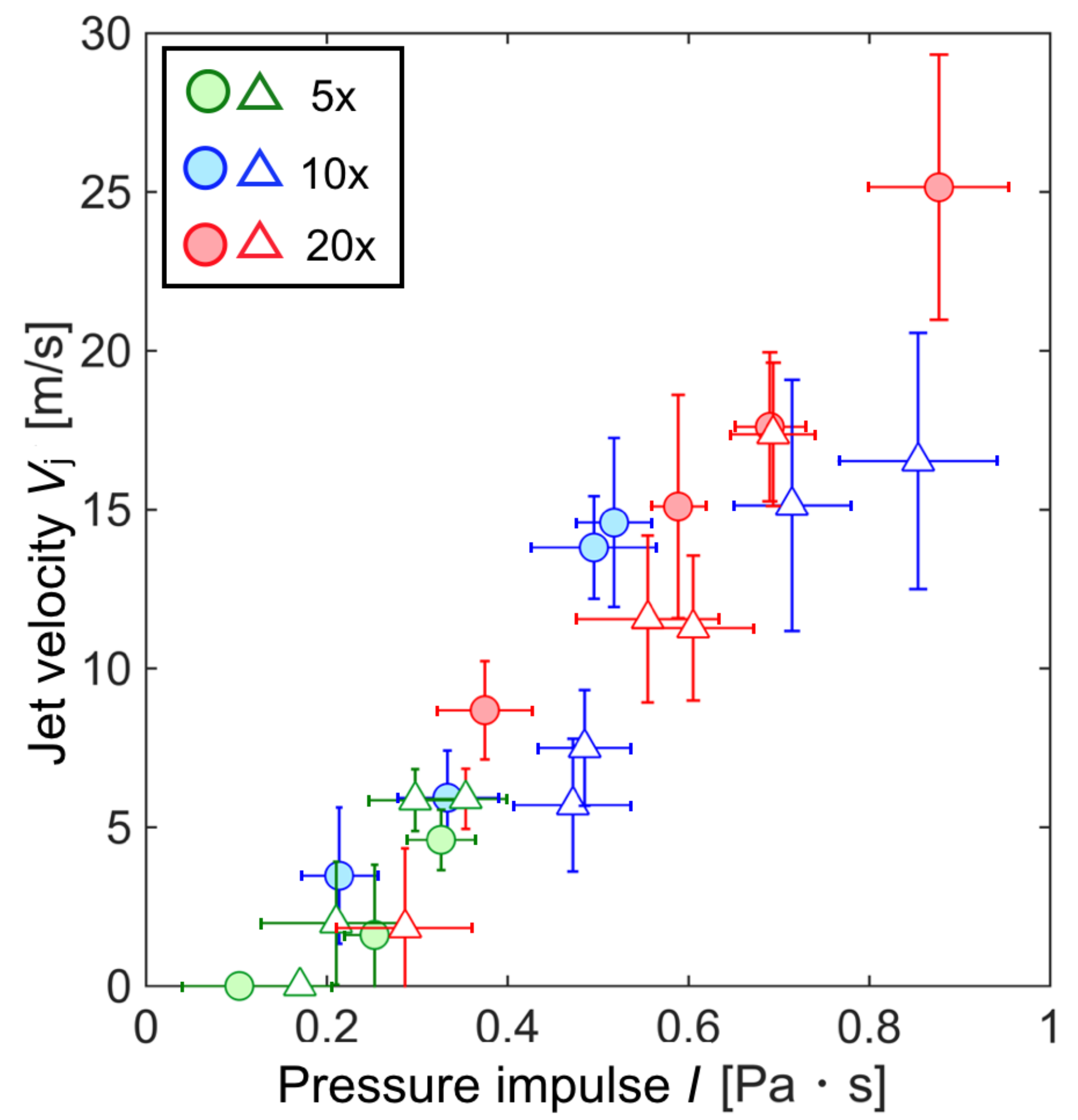}
\caption{
Microjet velocity vs. pressure impulse. 
Circle markers: the glass tube and the hydrophone are arranged on $\theta$ = 0$^\circ$ and $\theta$ = 90$^\circ$, respectively (see Fig.\ref{SetupB}(a)). 
Triangle markers: the glass tube and the hydrophone are arranged on $\theta$ = 90$^\circ$ and $\theta$ = 0$^\circ$, respectively (see Fig.\ref{SetupB}(b)). 
Green, blue and red plots designate experiments with a 5$\times$, 10$\times$, 20$\times$ objective microscope lens, respectively. 
The error bars correspond to the standard deviation calculated from 5 trials. 
}
\label{Impulse}       
\end{figure}

\begin{figure}[h!]
\includegraphics[width=0.95\columnwidth]{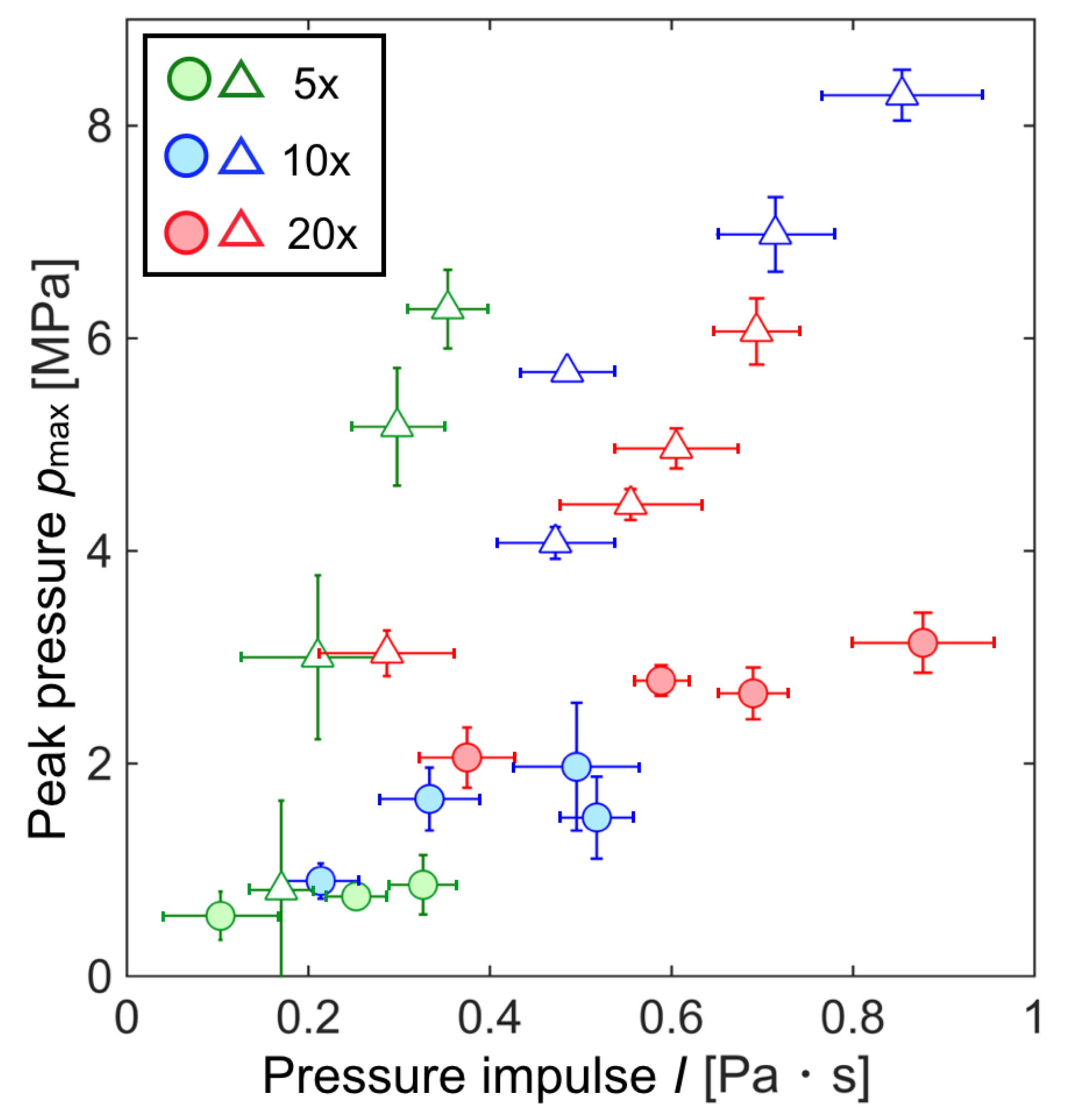}
\caption{
Peak pressure vs. pressure impulse. 
Markers and colors mean the same as shown in Fig.\ref{Impulse}. 
}
\label{Peak}       
\end{figure}

First, we consider the effect of the shock wave pressure impulse on \tr{microjet velocity.}
The pressure impulse is given by the following equation:

\begin{equation}
\label{impulse}
I=\int { pdt }   
\end{equation}

\noindent
Here, $I$ indicates the pressure impulse, $t$ indicates the time that the pressure impulse acts on the fluid, and $p$ indicates the pressure that acts on the fluid in a short period of time. 
Under the experimental conditions of this study, the time at which the shock wave acts on the interface is between 2-5 $\mu$s after the laser pulse is emitted (see Fig. \ref{Wave1}). 
Fig. \ref{Impulse} shows the jet velocity vs. the pressure impulse, where the range of laser energy is 3-10 mJ. 
The horizontal axis of the graph shows the pressure impulse obtained using the hydrophone, and the vertical axis shows the velocity of the jet generated in the tube. 
The plots represent mean values obtained after conducting the test five times, and the error bars indicate the corresponding standard deviations. 
When the pressure impulse is below 0.2 Pa$\cdot$s, the jet velocity is zero. 
When the pressure impulse is above 0.2 Pa$\cdot$s, the jet velocity is roughly proportional to the pressure impulse. 
In addition, the jet velocity is almost constant regardless of the type of objective lens and the orientation of the equipment. 
The pressure impulse differs greatly for each objective lens, even though the range of laser energy is the same. 
The reason for this is considered to be the laser focusing shape\cite{vogel1999influence}.

Fig. \ref{Peak} shows the peak pressure vs. the pressure impulse measured using the hydrophone. 
The peak pressure and pressure impulse are measured simultaneously using one hydrophone. 
The peak pressure is proportional to the pressure impulse, but the gradient strongly depends on the propagation direction. 
The peak pressure for $\theta$ = 90$^\circ$ is approximately three times higher than that for $\theta$ = 0$^\circ$. 
If the jet velocity was dependent on the peak pressure, the jet velocities in Fig. \ref{Impulse} would vary according to the direction of propagation. 
However, since this is not the case, the jet velocity is considered to be dependent on the pressure impulse, rather than the peak pressure.
\tr{
The timescale for jet formation $\tau$$\sim$$r$/$V_{j}$ is approximate 25 $\mu$s ($r$=250 $\mu$s, $V_{j}\approx$10 m/s), which is much larger than pressure duration of the shock wave ($\leq$2 $\mu$s, see Fig.\ref{Wave1}).
Therefore the entire pressure impulse contributes to jet formation, and determines microjet velocity. 
Note that the Laplace pressure ($p=2\sigma/R$) on free-surface is approximately 0.3 kPa ($\sigma \sim$70 mN/m, surface curvature $R$$\sim$250 $\mu$m), which is much lower than the shock pressure ($\sim O$(1) MPa).
It indicates that the surface tension does not hinder the jet formation for both $\theta$ = 0$^\circ$ and $\theta$ = 90$^\circ$.
}

\section{Results and discussion \\ (L-shaped microchannel experiments)}
\label{Sec:Result_B}


\subsection{Probability of cavitation}
\label{SubSec:B1}

\begin{figure}[t]
\includegraphics[width=1.0\columnwidth]{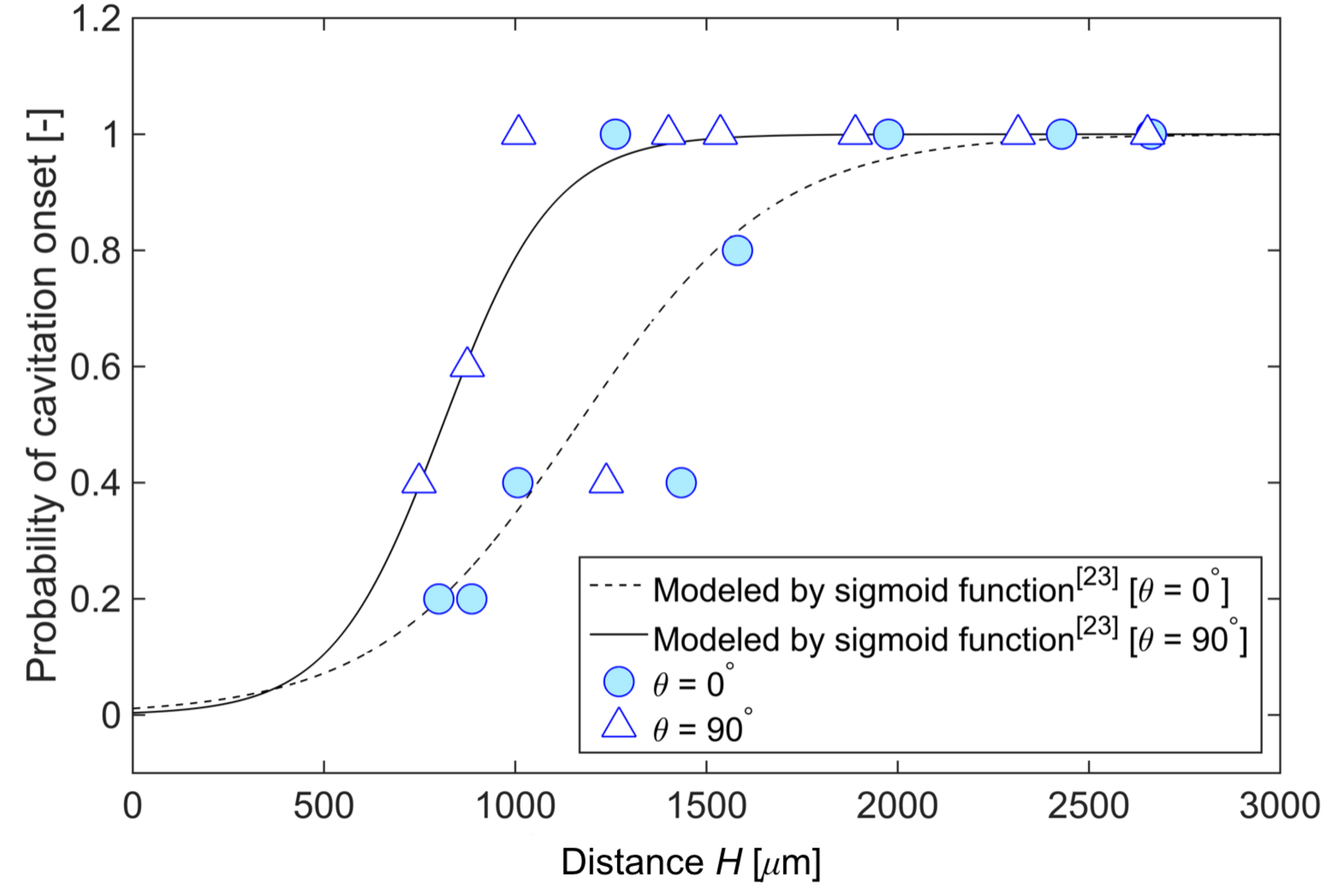}
\caption{
Probability of cavitation as a function of distance from a laser-induced bubble to meniscus. 
The circles and triangles are the probabilities obtained for $\theta$=0$^\circ$ and $\theta$=90$^\circ$, respectively. 
The black line and dotted line are fits using a sigmoid function.
}
\label{Cav}       
\end{figure}

This section describes the effect of a shock wave with an non-spherically-symmetric pressure distribution on the probability of cavitation inside an L-shaped \tb{microchannel}. 
In the water tank experiments, when the distance from the laser-induced bubble to the air-liquid interface was 5 mm, the lowest peak pressure of the underwater shock wave was approximately 0.5 MPa (Fig. \ref{Impulse}). 
If the shock wave traveling inside the L-shaped tube is comparable to that in the water tank, the expansion wave occurring at the air-liquid interface inside the L-shaped tube is considered to be smaller than the saturated vapor pressure of water ($\approx$ 2 kPa). 
Therefore, the expansion wave is considered to have a negative pressure that can cause cavitation in the liquid. 
Fig. \ref{Cav} shows the probability of cavitation occurring in the L-shaped \tb{microchannel}. 
The horizontal axis is the distance $H$ from the laser-induced bubble to the air-liquid interface, and the vertical axis is the probability of cavitation. 
Each plot represents the probability from five trials. 
In this case, the laser energy is constant at 650 $\mu$J. 
As $H$ increases, the amount of bubble nuclei is considered to increase with the volume of fluid. 
Therefore, the probability of cavitation increases with $H$\cite{kiyama2016effects}. 
The probability of cavitation is approximated using the least-squares method using the sigmoid function [Prob. = 1$\cdot$[1+exp$\langle$-$\{$(x-a)/b$\}$$\rangle$]$^{-1}$]\cite{maxwell2011cavitation}. 
The solid line (a = 811, b = 145) indicates the approximate curve for $\theta$ = 90$^\circ$, and the dotted line (a = 1164, b = 260) indicates the approximate curve for $\theta$ = 0$^\circ$. 
When $H\leq$ 2000, the probability of cavitation at $\theta$ = 90$^\circ$ is greater than at $\theta$ = 0$^\circ$. 
When $H$$>$2000, the probability of cavitation at $\theta$ = 90$^\circ$ and at $\theta$ = 0$^\circ$ is 1.

Cavitation can occur when the pressure around the bubble nucleus is lower than the vapor pressure. 
A large negative pressure increases the probability of cavitation\cite{maxwell2011cavitation}. 
The magnitude of the negative pressure is considered to be proportional to the magnitude of the shock wave before it is reflected at the air-liquid interface (Fig. \ref{Wave1}). 
Therefore, the magnitude of the peak value of negative pressure of the underwater shock wave propagating at $\theta$ = 90$^\circ$ is considered to be greater than that at $\theta$ = 0$^\circ$.

For $\theta$ = 90$^\circ$, the air-liquid interface is positioned perpendicular to the laser irradiation direction. 
As a result, for $\theta$ = 90$^\circ$, a shock wave with a high peak pressure propagates, and the probability of cavitation is considered to increase. 
Meanwhile, for $\theta$ = 0$^\circ$, the air-liquid interface is positioned in the laser irradiation direction. 
As a result, for $\theta$ = 0$^\circ$, a shock wave with a low peak pressure propagates, and the probability of cavitation is considered to decrease. 
The above results suggest that the probability of cavitation is affected by the magnitude of the peak pressure of the shock wave.

\tr{
Here we discuss a possible scenario based on the theory of bubble \tg{dynamics}.
The lifetime of cavitation bubble\cite{delale2012bubble} in this study is approximate 20-30 $\mu$s, much larger than pressure duration ($\sim O$(1) $\mu$s).
Nevertheless, pressure impulse does not play any dominant role.
It is likely that the force balance in rapid expansion of a bubble is important.
The motion of a homogeneous bubble is described the Rayleigh-Plesset equation\cite{Dugue1992dynamic}.
This equation indicates that the expansion of a small bubble (i.e., cavitation nucleus) requires large pressure difference surpassing the Laplace pressure (= $2\sigma/R$) across the bubble surface.
Our experimental results suggest that the Laplace pressure against bubble expansion is $\sim O$(1) MPa, much larger than that in the jet formation ($\S$\ref{SubSec:A2}). 
Unless the pressure difference exceeds the threshold, the bubbles do not expand.
Thus, the pressure impulse should not be considered as the criterion of cavitation onset.
}

\subsection{Changes in microjet velocity with presence/absence of cavitation}
\label{SubSec:B2}

\begin{figure}[t!]
\includegraphics[width=0.85\columnwidth]{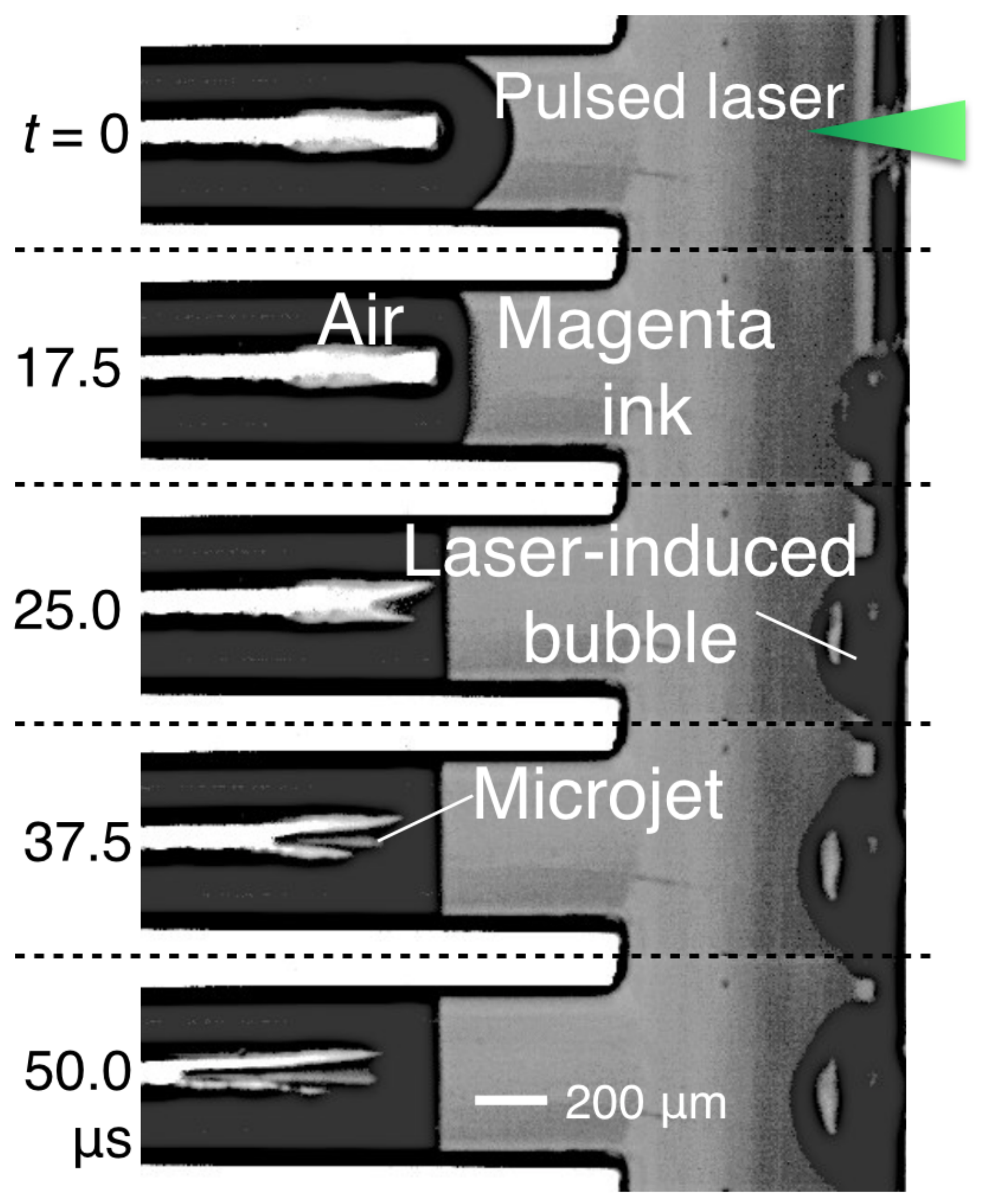}
\caption{
Photograph of the microjet generated during experiments in the L-shaped microchannel for $\theta$ = 0$^\circ$.  
$H$ = 1080 $\mu$m and the frame interval is 17.5 $\mu$s. 
The laser is incident from the right side and the laser energy is 185 $\mu$J.
}
\label{SnapA}       
\end{figure}

\begin{figure}[t!]
\includegraphics[width=1.0\columnwidth]{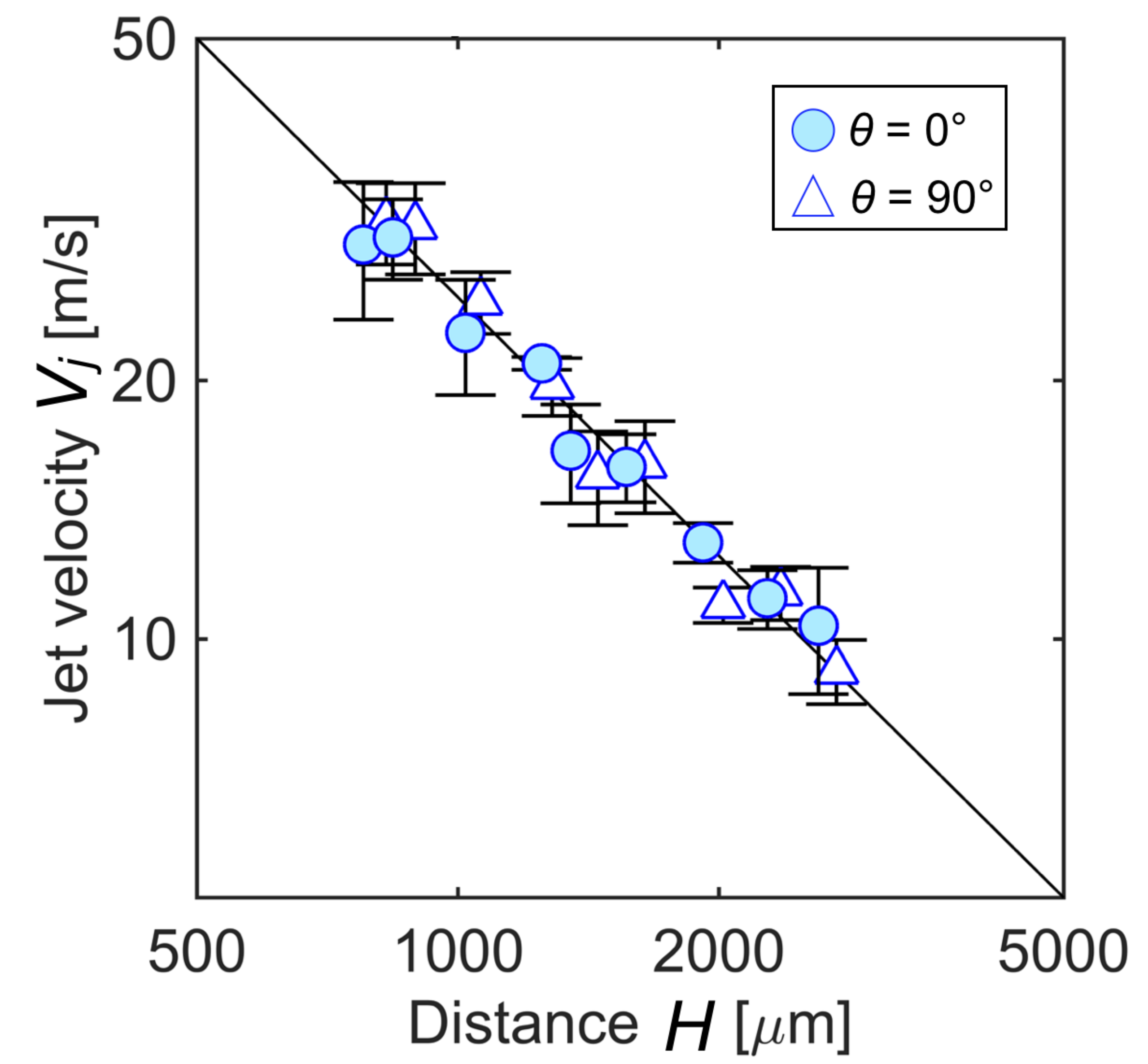}
\caption{
Jet velocity as a function of distance from laser-induced bubble to the air-liquid interface. 
The laser is incident from the right side and the laser energy is 185 $\mu$J.
The circles and triangles designate results for $\theta$ = 0$^\circ$  and $\theta$ = 90$^\circ$, respectively. 
The error bars are the standard deviation.
}
\label{Velocity_A}       
\end{figure}

This section shows the results of measuring jet velocity focusing on the presence or absence of cavitation in the L-shaped microchannel experiments. 
First, we look at microjets when the laser energy is relatively low and cavitation does not occur. 
The jet velocity is not subject to the action of cavitation, so it is expected to depend on the magnitude of pressure impulse and not to change with the laser irradiation direction. 
Fig. \ref{SnapA} shows snapshots taken during microjet formation for $\theta$ = 0$^\circ$. 
In the initial state ($t$ = 0), the air-liquid interface has a concave shape. 
A laser-induced bubble occurs ($t$ = 17.5 $\mu$s) at the wall through which the laser passes. 
The air-liquid interface displaces immediately after laser irradiation, and a focused microjet forms. 
Due to the limitations of the photographic equipment, it is difficult to photograph the microjet and the shock wave simultaneously. 
Because a laser-induced bubble occurs, it is thought that the shock wave develops \trr{in association with} rapid expansion of the bubble, and the microjet forms as a result from the air-liquid interface ($t$ = 25.0 $\mu$s). 
Next, Fig. \ref{Velocity_A} shows the jet velocity $V_{j}$ for various values of $H$, in the case of $\theta$ = 0$^\circ$ and $\theta$ = 90$^\circ$. 
The vertical and horizontal axes are logarithmic. 
The laser energy is constant at 185 $\mu$J. 
The plots represent mean values obtained after conducting the test five times, and the error bars indicate the standard deviations. 
The solid line shows the gradient of the inversely proportional relationship. 
For $\theta$ = 0$^\circ$, the contact angle $\theta$ between the air and the liquid at the air-liquid interface changes with $H$ due to the effect of gravity, so the values of $V_{j}$ take account of changes in the contact angle between the air and liquid\cite{tagawa2012highly}. 
The jet velocity $V_{j}$ decreases with increasing $H$. 
It is thought that $V_{j}$ is inversely proportional to the distance because the shock wave pressure decays in inverse proportion to the propagation distance\cite{vogel1996shock}. 
In addition, the jet velocity for $\theta$ = 0$^\circ$ is approximately equal to for $\theta$ = 90$^\circ$. 
The pressure impulse propagating to the air-liquid interface was equal in each setup, so the jet velocity is considered to be roughly constant regardless of direction, as expected.

\begin{figure}[t]
\includegraphics[width=1.0\columnwidth]{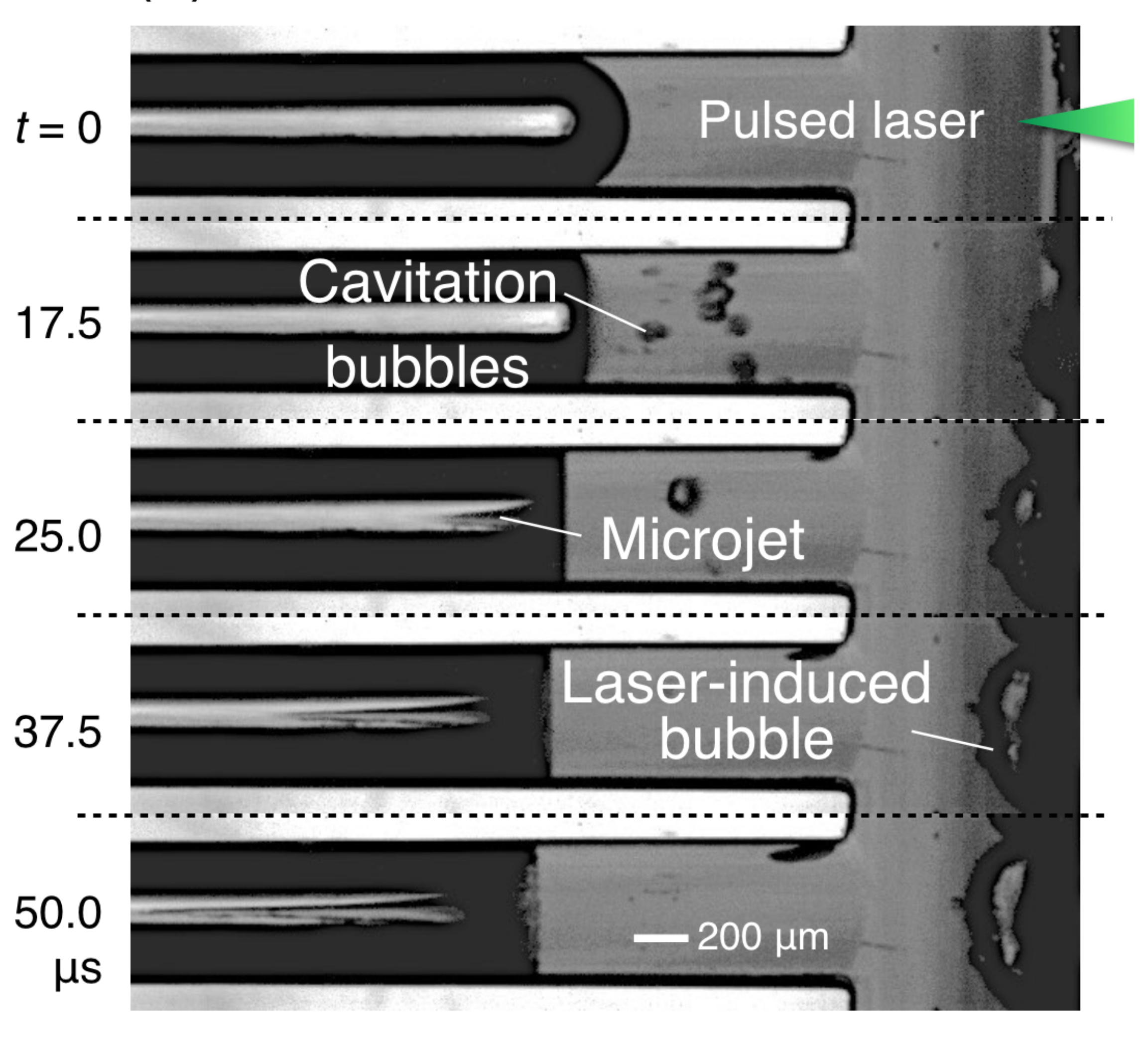}
\caption{
Photograph  of  the  microjet  generated  at  $\theta$ = 0$^\circ$  with the cavitation bubbles. 
$H$ = 1080 $\mu$m and the frame interval is 17.5 $\mu$s. 
The laser is incident from the right side and the laser energy is 650 $\mu$J.
}
\label{SnapB}       
\end{figure}

\begin{figure}[t]
\includegraphics[width=1.0\columnwidth]{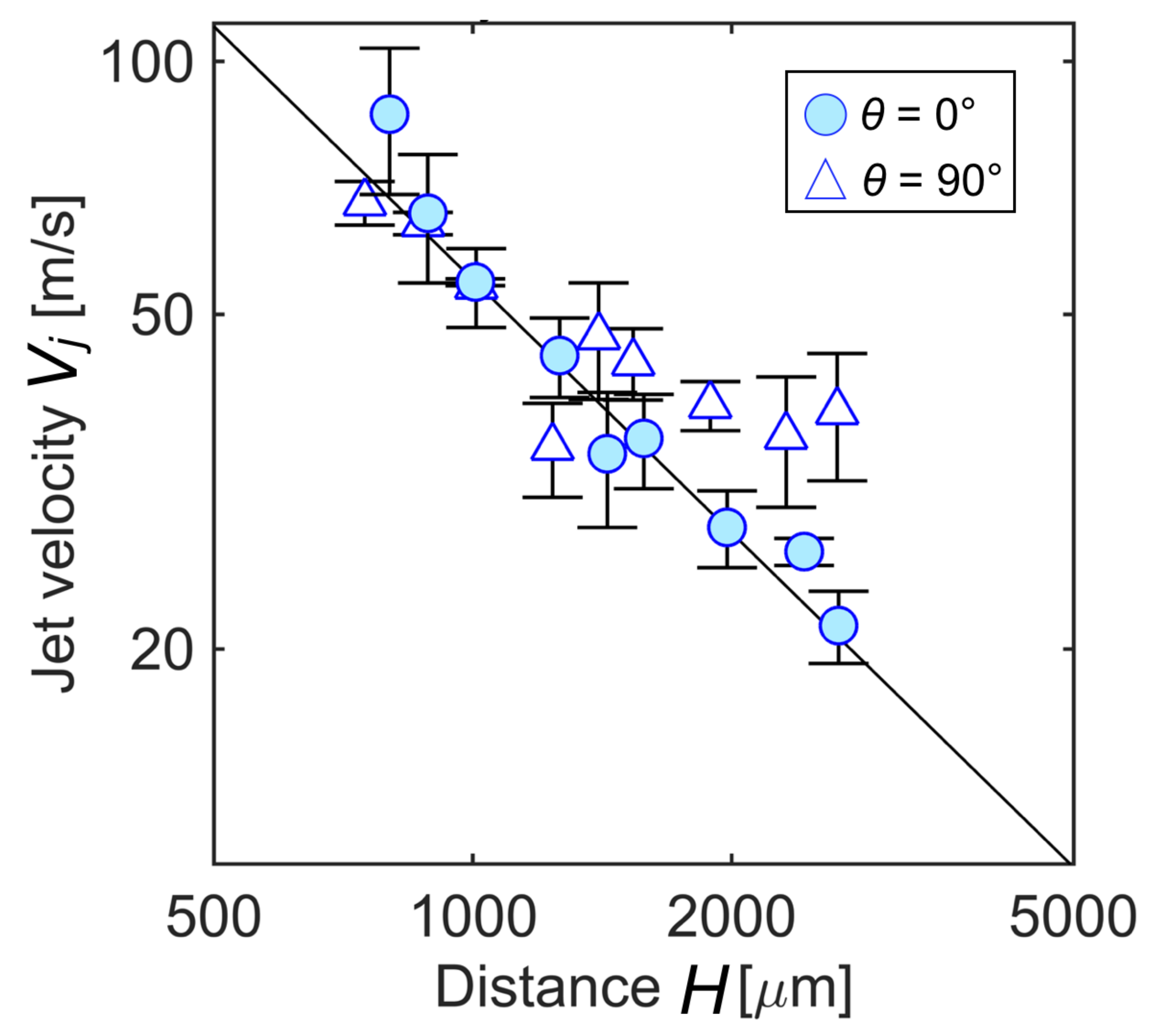}
\caption{
Jet velocity as a function of distance from laser-induced bubble to the air-liquid interface. 
The laser is incident from the right side and the laser energy is 650 $\mu$J.
The circles and triangles designate results for $\theta$ = 0$^\circ$  and $\theta$ = 90$^\circ$, respectively. 
The error bars are the standard deviation. 
}
\label{Velocity_B}       
\end{figure}

Next, we look at microjets when the laser energy is relatively high and cavitation occurs. 
Fig. \ref{SnapB} shows snapshots taken during microjet formation for $\theta$ = 0$^\circ$. 
The distance $H$ is 2000 $\mu$m. When $t$ = 17.5 $\mu$s, cavitation occurs near the air-liquid interface, which is thought to have been generated by an expansion wave reflected by the shock wave at the air-liquid interface. 
In general, cavitation can occur when the pressure in a liquid at a bubble nucleus is lower than the vapor pressure of the liquid. 
The peak pressure of the shock wave increases with the laser energy. 
Therefore, the negative pressure of the expansion wave produced by reflection at the air-liquid interface increased and cavitation occurred. 
Fig. \ref{Velocity_B} shows the relationship between $H$ and $V_{j}$. 
These jet velocities were measured at the same time as the probability of cavitation (Fig. \ref{Cav}). 
The laser energy is constant at 650 $\mu$J. 
Cavitation occurred in most cases when the microjets measured in Fig.\ref{Velocity_B} formed. 
When $H$$\leq$1500, the relationship between $H$ and $V_{j}$ is inversely proportional, and this agrees well with the trend shown when the laser energy is weak (Fig. \ref{Velocity_A}). 
When $H$$>$1500, the results for $\theta$ = 0$^\circ$ approximate the solid line indicating an inversely proportional relationship. 
On the other hand, in the case of $\theta$ = 90$^\circ$, the velocities are higher than the solid line indicating an inversely proportional relationship. 
When there was increased velocity, cavitation always occurred. 

Kiyama \etal\cite{kiyama2016effects} identified the possibility of relaxation of negative pressure in the liquid due to the occurrence of cavitation as a cause of increased velocity. 
The larger the maximum cavitation volume, the greater the amount of relaxation of negative pressure due to cavitation. 
Under the experimental conditions used in this study, there were an average of two images that captured cavitation per trial, so it is difficult to determine maximum cavitation volume. 
However, in light of the fact that the probability of cavitation differs between $\theta$ = 0$^\circ$ and $\theta$ = 90$^\circ$, the maximum cavitation volume in either direction is also expected to differ in a similar manner. 
The jet velocity for $\theta$ = 90$^\circ$ is considered to have been higher than for $\theta$ = 0$^\circ$, because the maximum cavitation volume for $\theta$ = 90$^\circ$ was greater than for $\theta$ = 0$^\circ$. 
Based on the above findings the jet velocity is considered to change according to the cavitation volume, and the jet velocity at the onset of cavitation is considered to be affected by the peak pressure.

\trr{
Above findings might contribute to improve controllability of microjet velocity.
When $H$ is relatively small, cavitation does not occur in a microchannel. 
Thus the jet velocity is constant irrespective to the direction of a laser irradiation, leading to the increase in freedom for designing microjet generators.
When $H$ and the laser energy are relatively large, cavitation occurs.
Although it can cause the cavitation-assisted increment of microjet velocity, the controllability of jet velocity gets worse.
Our experimental results (see Fig.\ref{Cav}) indicate that a suitable choice of the direction reduces cavitation onset, leading to high controllability of microjet velocity.
}

\section{Conclusion}
\label{Sec:Con}

In this paper, we experimentally investigated the effect of laser-induced underwater shock waves on high-speed microjet velocity and underwater cavitation, focusing on the pressure impulse and the peak pressure of the shock wave. 
We focused on a characteristic of non-spherically-symmetric shock waves: namely, a non-spherically symmetric peak pressure distribution and a spherically symmetric pressure impulse distribution. 
We made use of this characteristic to measure the pressure and jet velocity parallel and perpendicular to the direction of laser irradiation.

Using the first experimental setup, we investigated the effect of the shock wave pressure on the jet velocity in a water tank. 
We demonstrated that the jet velocity is dependent on the magnitude of the pressure impulse, regardless of the magnitude of the peak pressure. 
Using the second experimental setup, we investigated the effect of the shock wave pressure on cavitation in an L-shaped \tb{microchannel}. 
We showed that the probability of cavitation changes according to the magnitude of the peak pressure. 
The jet velocity at the onset of cavitation can be considered to be affected by the peak pressure.
\trr{In addition we discuss controllability of micojet velocity in a microchannel.
We suggest that our novel findings have potentials to improve flexibility of the design or controllability of microjet velocity.}

\begin{acknowledgements}
This work was supported by JSPS KAKENHI Grant Number 26709007.
\end{acknowledgements}

\bibliographystyle{unsrt} 
\bibliography{rsc}

\end{document}